\documentclass[12pt]{article}
\usepackage{amssymb}
\usepackage[english]{babel}
\usepackage{cite}

\newcommand{\nc}{\newcommand}

\newcommand{\be}{\begin{equation}}
\newcommand{\ee}{\end{equation}}
\newcommand{\bes}{\begin{equation*}}
\newcommand{\ees}{\end{equation*}}

\newcommand{\bea}{\begin{eqnarray}\displaystyle}
\newcommand{\eea}{\end{eqnarray}}

\nc{\NN}{{\cal N}}
\nc{\OO}{{\cal O}}
\nc{\KK}{{\cal K}}
\nc{\PP}{{\cal P}}
\nc{\UU}{{\cal U}}
\nc{\TO}{{\cal T}}
\nc{\KB}{{\cal K}}
\nc{\AN}{{\cal A}}
\nc{\MS}{{\cal M}}
\nc{\CC}{{\cal C}}
\nc{\DD}{{\cal D}}
\nc{\HH}{{\cal H}}
\nc{\GG}{{\cal G}}
\nc{\YY}{{\cal Y}}
\nc{\VV}{{\cal V}}
\nc{\II}{{\cal I}}

\nc{\BZ}{{\cal Z}}

\nc{\CY}{C{\rm{Y}_3}}
\nc{\ox}{\otimes}
\nc{\x}{\times}
\nc{\w}{\wedge}
\nc{\W}{\bigwedge}
\nc{\p}{\partial}
\nc{\bp}{\bar{\partial}}
\nc{\pbar}{\bar{\partial}}
\nc{\MM}{{\sf M}}
\nc{\wt}{\widetilde}
\nc{\wh}{\widehat}
\nc{\vp}{\varphi}
\nc{\ep}{\epsilon}
\nc{\vep}{\varepsilon}
\nc{\vtheta}{\vartheta}

\nc{\RR}{\mathbf{R}}
\nc{\RE}{{\cal R}}
\nc{\Z}{\mathbf{Z}}
\nc{\ZZ}{\mathbf{Z_2}}
\nc{\RP}{\mathbb{R}\mathbb{P}}
\nc{\cplx}{\mathbf{C}} % Complex Plane = Cplx Plane
\nc{\one}{{\mathbf{1}}}

\usepackage{babel}
\makeatother

\begin{document}

\rightline{HIP-2008-29/TH}

%\leftline\today\rightline\currenttime

\vskip 1cm \centerline{\large {\bf Dynamical renormalization group methods}}
\centerline{\large {\bf in theory of eternal inflation}}
\vskip 1cm
\renewcommand{\thefootnote}{\fnsymbol{footnote}}
\centerline{{\bf Dmitry Podolsky,\footnote{dmitry.podolsky@helsinki.fi; \emph{URL}: www.nonequilibrium.net}
 }}
\vskip .5cm \centerline{\it Helsinki Institute of Physics} \centerline{\it
P.O.Box 64, FIN-00014 University of Helsinki, Finland}

\begin{abstract}
Dynamics of eternal inflation on the landscape admits description in terms of the Martin-Siggia-Rose (MSR) effective field theory that is in one-to-one correspondence with vacuum dynamics equations.
On those sectors of the landscape, where transport properties of the probability measure for eternal inflation are important, renormalization group fixed points of the MSR effective action determine late time behavior of the probability measure. I argue that these RG fixed points may be relevant for the solution of the gauge invariance problem for eternal inflation.
\end{abstract}

\setcounter{footnote}{0}
\renewcommand{\thefootnote}{\arabic{footnote}}

\newpage

Consider a single field inflation driven by the inflaton field $\phi$
with potential $V(\phi)$. During one Hubble time $\Delta t\sim H^{-1}$
the value of the inflaton field changes by
%%%%%%%%%%%%%%%%%%%%%%%
\begin{equation}
\Delta\phi\sim\dot{\phi}\Delta t\sim\frac{1}{3H^{2}}\frac{\partial V}{\partial\phi}\sim
\frac{M_{P}^{2}}{8\pi V}\frac{\partial V}{\partial\phi}
\label{eq:Deltaphiclass}
\end{equation}
%%%%%%%%%%%%%%%%%%%%%%%
under the action of the classical force $-\frac{\partial V}{\partial\phi}$,
and superhorizon fluctuations $\delta\phi$ of the inflaton are generated
with characteristic amplitude
%%%%%%%%%%%%%%%%%%%%%%%
\begin{equation}
|\delta\phi|\sim\frac{H}{2\pi}\sim\sqrt{\frac{2V}{3\pi M_{P}^{2}}}.
\label{eq:Deltaphiquant}
\end{equation}
%%%%%%%%%%%%%%%%%%%%%%%
When
%%%%%%%%%%%%%%%%%%%%%%%
\begin{equation}
|\delta\phi|\gg\Delta\phi,
\label{eq:EternalInfCondition}
\end{equation}
%%%%%%%%%%%%%%%%%%%%%%%
the ``quasiclassical'' picture of the inflaton slowly rolling towards
the minimum of its potential $\phi=\phi_{0}$, $V'(\phi_{0})=0$ breaks
down, and we say that inflation enters eternal selfreproducing regime
\cite{LindeEternalInflation}.

The physical picture emerging in this regime can be well understood
in stochastic formalism \cite{StochasticInflation,NakaoNambuSasaki}. As we know,
the causal structure of eternally inflating Universe can be represented
as an infinite set of Hubble patches with values of $\phi$ and the
effective cosmological constant $\frac{V(\phi)}{M_{P}^{2}}$ being
random, stochastically distributed among the patches. The corresponding
probability distribution $P(\phi,t)$ (its physical meaning is the
probability density to measure a given value of the inflaton in a
given Hubble patch) is the solution of the Fokker-Planck equation
%%%%%%%%%%%%%%%%%%%%%%%
\begin{equation}
\frac{\partial P(\phi,t)}{\partial t}=\hat{H}P=\frac{H_{0}^{3}}{8\pi^{2}}\frac{\partial^{2}P}{\partial\phi^{2}}+\frac{1}{3H_{0}}
\frac{\partial}{\partial\phi}\left(\frac{\partial V}{\partial\phi}P\right)
\label{eq:fokkerplanck}
\end{equation}
%%%%%%%%%%%%%%%%%%%%%%%
(for simplicity we consider $V(\phi)=V_{0}+\delta V(\phi)$, where
$|\delta V(\phi)|\ll V_{0}$; the Eq. (\ref{eq:fokkerplanck}) can
be relatively easily generalized to the case of arbitrary potential
$V(\phi)$ \cite{StochasticInflation}), and IR contributions to
the correlation functions of the inflaton $\phi$ (including arbitrary loop corrections) can be simply calculated
\cite{EnqNurPodRig} by taking integrals
%%%%%%%%%%%%%%%%%%%%%%%
\begin{equation}
\langle\phi^{n}\rangle=\int d\phi\phi^{n}P(\phi,t).
\end{equation}
%%%%%%%%%%%%%%%%%%%%%%%
The general solution of the Eq. (\ref{eq:fokkerplanck}) is represented
as the expansion
%%%%%%%%%%%%%%%%%%%%%%%
\begin{equation}
P(\phi,t)=\sum_{n=0}^{+\infty}\psi_{n}(\phi)e^{-E_{n}t}
\label{eq:expansion}
\end{equation}
%%%%%%%%%%%%%%%%%%%%%%%
over the eigenfunctions of the operator $\hat{H}$. All eigenvalues
$E_{n}$ are non-negative since the {}``Hamiltonian'' $\hat{H}$
is always supersymmetric. Moreover, if the normalization integral
%%%%%%%%%%%%%%%%%%%%%%%
\begin{equation}
\int d\phi e^{-\frac{8\pi^{2}\delta V(\phi)}{3H^{4}}}
\label{eq:NormalizationIntegral}
\end{equation}
%%%%%%%%%%%%%%%%%%%%%%%
converges, the lowest eigenvalue of $\hat{H}$ is $E_{0}=0$, so that
at $t\to\infty$ the distribution $P(\phi,t)$ approaches the time independent
asymptotics given by
%%%%%%%%%%%%%%%%%%%%%%%
\begin{equation}
P_{{\rm asymp}}(\phi)\sim\exp\left(-\frac{8\pi^{2}\delta V(\phi)}{3H_{0}^{4}}\right).
\label{eq:Asymp}
\end{equation}
%%%%%%%%%%%%%%%%%%%%%%%
This expression is known as the Hartle-Hawking wavefunction of the
Universe \cite{HartleHawking}.\footnote{It is interesting to note that the Linde-Vilenkin wavefunction of
the Universe \cite{LindeVilenkin} is \emph{not} a solution of the
Eq. (\ref{eq:fokkerplanck}), so the stochastic formalism allows us
to unambiguously resolve the long debate between advocates of the
Hartle-Hawking and Linde-Vilenkin wavefunctions.}

Discussing physics of eternal inflation, one was usually interested
in this time-independent asymptotics because of two reasons: 1) time
scale $\delta t_{{\rm asymp}}$ necessary for $P$ to reach the
asymptotics (\ref{eq:Asymp}) is rather short in simple inflationary
models: for example, for $\delta V(\phi)=\frac{1}{2}m^{2}\phi^{2}$
we find $\delta t_{{\rm asymp}}\sim E_{1}^{-1}=\frac{3H_{0}}{m^{2}}$,
for $\delta V(\phi)=\frac{1}{4}\lambda\phi^{4}$ --- $\delta t_{{\rm asymp}}\sim\lambda^{-1/2}H_{0}^{-1}$,
etc., 2) the asymptotic distribution (\ref{eq:Asymp}) does not depend
on initial conditions for eternal inflation, while the contributions
of higher eigenstates into the expansion (\ref{eq:expansion}) do.

However, in many physically interesting situations such as for eternal
inflation on the string theory landscape \cite{PodolskyEnqvist,PodolskyJokelaMajumder}
the time scale $\delta t_{{\rm asymp}}$ is extremely long --- very
roughly one can estimate it as $\delta t_{{\rm asymp}}\sim E_{1}^{-1}\sim N\epsilon_{1}^{-1}$,
where $N$ is the number of vacua on the landscape and $\epsilon_{1}$
is the eigenvalue corresponding to first excited state in the problem
with a single dS vacuum. Also, often the time independent asymptotics
(\ref{eq:Asymp}) may not exist at all.\footnote{This is the case for chaotic inflationary models with potentials $V(\phi)=\lambda_{n}\phi^{n}$, when the normalization integral (\ref{eq:NormalizationIntegral})
for the Hartle-Hawking wavefunction diverges at small $\phi$ \cite{StochasticInflation}.}
In such situation, since it is clearly impossible to solve the Fokker-Planck
equation (\ref{eq:fokkerplanck}) analytically for arbitrary $V(\phi)$,
one can use various approximation schemes to analyze behavior of the
probability distribution $P(\phi,t)$ at intermediate time scales
$M_{P}^{-1}\ll t\ll E_{1}^{-1}$. One such scheme is based on the
method of dynamical renormalization group analysis \cite{DynRenorm},
and it works especially well when transport properties of the landscape encoded in the Eq.
(\ref{eq:fokkerplanck}) are important, such as for an infinite landscape
with effective potential/Hawking-Moss/Coleman-de Luccia tunneling rates not
growing at infinity \cite{PodolskyEnqvist,PodolskyJokelaMajumder}.

Very roughly, the idea is the following. Let us consider a $n$-point
correlation function of the inflaton in the limit when all points
are taken within a single Hubble patch. It can be represented in the
following functional form
%%%%%%%%%%%%%%%%%%%%%%%
\bea\label{eq:FuncForm}
\langle\phi^{n}\rangle & = & \int d\phi\phi^{n}P(\phi,t)=\int d\phi\phi^{n}\int{\cal D}P\delta\left(\frac{\partial P}{\partial t}-\hat{H}P\right)\\
& = & \int d\phi\phi^{n}\int{\cal D}P{\cal D}\bar{P}\exp\left(i\int dtd\phi\bar{P}\left(\partial_{t}-\hat{H}\right)P\right)
\eea
%%%%%%%%%%%%%%%%%%%%%%%
since the probability distribution $P(\phi,t)$ satisfies the Fokker-Planck
equation (\ref{eq:fokkerplanck}). In the limit $t\to\infty$ (but
$t$ sufficiently small so that $|P(\phi,t)-P_{{\rm asymp}}(\phi)|\gg P_{{\rm asymp}}(\phi)$)
only renormalizable terms survive in the ``effective action''
%%%%%%%%%%%%%%%%%%%%%%%
\begin{equation}
S_{{\rm eff}}=\int dtd\phi\bar{P}\left(\partial_{t}-\hat{H}\right)P,
\label{eq:EffectiveAction1}
\end{equation}
%%%%%%%%%%%%%%%%%%%%%%%
and scaling analysis of the latter allows us to learn many things
about long time behavior of correlation functions $\langle\phi^{n}\rangle$.
In particular, it turns out that this behavior
is determined by fixed points of the RG flow for the effective action
(\ref{eq:EffectiveAction1}). Let me now present two examples of landscapes
with RG fixed points. After that I will discuss how RG fixed points
for the effective action (\ref{eq:EffectiveAction1}) may be relevant
for the resolution of the famous gauge invariance problem for eternal
inflation \cite{GaugeInvariance}.

First, let us consider stochastic inflation in the potential with a weak \emph{disordered}
component $V(\phi)=V_{0}+\delta V(\phi)$, $|\delta V|\ll V_{0}$
\cite{PodolskyEnqvist}. In the limit $|\delta V|\to0$ the solution
of the Fokker-Planck equation (\ref{eq:fokkerplanck}) is given by
%%%%%%%%%%%%%%%%%%%%%%%
\begin{equation}
P(\phi,t)\sim\frac{1}{\sqrt{t}}e^{-\frac{2\pi^{2}(\phi-\phi_{0})^{2}}{H_{0}^{3}t}},
\label{eq:Gaussian}
\end{equation}
%%%%%%%%%%%%%%%%%%%%%%%
and we want to understand how the linear diffusion of the probability
distribution $P(\phi,t)$ is affected by weak disorder $\delta V(\phi)$
in the limit $t\to\infty.$

For simplicity we take a gaussian distributed random potential with
the correlation properties
%%%%%%%%%%%%%%%%%%%%%%%
\bea\label{eq:PotentialCorrProperties}
\langle\delta V(\phi)\rangle=0,& &\langle\delta V(\phi)\delta V(\phi')\rangle=\Delta\delta(\phi-\phi')
\eea
%%%%%%%%%%%%%%%%%%%%%%%
where the averaging $\langle\cdots\rangle$ is taken over the ensemble
of potentials $\delta V$, or, what is equivalent, over intervals
$(\phi,\phi+\delta\phi)$ of one particular realization of $\delta V$
in this ensemble.\footnote{Naively, it seems that the gaussian ensemble (\ref{eq:PotentialCorrProperties})
is not well defined, because it contains configurations corresponding
to negative values of the full potential $V(\phi)$. However, if one
changes the correlation properties (\ref{eq:PotentialCorrProperties})
accordingly to resolve this issue, one finds \cite{PodolskyJokelaMajumder}
that the change does not influence the long time behavior of the probability
distribution $P(\phi,t)$ since it introduces irrelevant terms into
effective action (\ref{eq:EffectiveAction1}).}

To determine how weak disorder affects dynamics of $P(\phi,t)$, one
can integrate it out in the ``partition function'' (\ref{eq:FuncForm})
and perform scaling analysis of the resulting effective action. One
has
%%%%%%%%%%%%%%%%%%%%%%
\bea\label{eq:EffActRandomPot}
S_{{\rm eff}} & = & \int dtd\phi\left(\bar{P}\frac{\partial}{\partial t}P-\frac{H_{0}^{3}}{8\pi^{2}}\bar{P}\frac{\partial^{2}}{\partial\phi^{2}}P-\right.\\
& &\left.-\frac{\Delta}{2}\int dt'\frac{\partial}{\partial\phi}\left(P(\phi,t)\frac{\partial\bar{P}(\phi,t)}{\partial\phi}\right)
\frac{\partial}{\partial\phi}\left(P(\phi,t')\frac{\partial\bar{P}(\phi,t')}{\partial\phi}\right)\right)
\eea
%%%%%%%%%%%%%%%%%%%%%%
For qualitative understanding of physics it is actually sufficient
to use simple scaling arguments: from the quadratic part of the effective
action (\ref{eq:EffActRandomPot}), by multiplying $\phi$ by a factor
of $l$ we immediately find that the time scale $t$ scales as $l^{2}$,
the product $\bar{P}P$ --- as $l^{-1}$ and the disorder strength
$\Delta$ --- as $l^{-1}$ as well. Therefore, we conclude that weak disorder
of the random potential type (\ref{eq:PotentialCorrProperties}) is irrelevant for the long time behavior
of $P(\phi,t)$ and does not influence the linear diffusion law $\langle\phi^{2}\rangle\sim t$.
The proper renormalization group analysis of the theory (\ref{eq:EffActRandomPot})
confirms this conclusion \cite{PodolskyEnqvist,PodolskyJokelaMajumder}.

Since we are also interested to know the intermediate time scale behavior
of the probability distribution $P(\phi,t)$, we can try to figure
out how important are the excited states $\psi_{k}(\phi)$, $k>0$ present
in the expansion (\ref{eq:expansion}) for the dynamics
of $P(\phi,t)$. As it turns out \cite{PodolskyJokelaMajumder,PhononsLocalization},
for any given realization of the disorder all eigenstates are \emph{localized}
or peaked near the points $\phi=\phi_{i}$ corresponding to the deepest
minima of the full potential $V(\phi)$. The width of $\psi_{k}(\phi)$
near a localization center $\phi_{i}$ behaves as\[
\delta\psi_{k}\sim E_{k}^{-1}\sim k^{-2},\]
growing with decreasing $k$. Excited, localized, eigenstates dominate in the expansion (\ref{eq:expansion})
at small $t$, and one has
%%%%%%%%%%%%%%%%%%%%%%
\begin{equation}
\langle(\phi-\phi_{i})^{2}\rangle\sim t^{2},
\end{equation}
%%%%%%%%%%%%%%%%%%%%%%
so the diffusion is slowed down significantly. At later times, the
regime of diffusion becomes linear, as we have found above.

Let us now turn to a somewhat different case of stochastic inflation
in the presence of random force $-\frac{\partial V}{\partial\phi}$,
when the correlation properties of disorder are
%%%%%%%%%%%%%%%%%%%%%%
\bea\label{eq:CorrelationPropertiesRandomForce}
\left\langle \frac{\partial V}{\partial\phi}\right\rangle =0,& &\left\langle \frac{\partial V}{\partial\phi}(\phi)\frac{\partial V}{\partial\phi}(\phi')\right\rangle =\sigma\delta(\phi-\phi').
\eea
%%%%%%%%%%%%%%%%%%%%%%
After integrating disorder out we find the following effective action
%%%%%%%%%%%%%%%%%%%%%%
\bea\label{eq:EffActRandomForce}
S_{\rm eff} &= &\int dtd\phi\left(\bar{P}\frac{\partial}{\partial t}P-\frac{H_{0}^{3}}{8\pi^{2}}\bar{P}\frac{\partial^{2}}{\partial\phi^{2}}P-\right.\\
& & \left.-\frac{\sigma}{2}\int dt'P(\phi,t)\frac{\partial\bar{P}(\phi,t)}{\partial\phi}P(\phi,t')\frac{\partial\bar{P}(\phi,t')}{\partial\phi}\right).
\eea
%%%%%%%%%%%%%%%%%%%%%%
Simple scaling analysis leads now to the conclusion that $\sigma$
scales as $l$, so that the disorder (\ref{eq:CorrelationPropertiesRandomForce})
is important for the long time dynamics of the probability distribution
$P(\phi,t)$. More accurate renormalization group analysis shows that
the RG fixed point $\sigma=\sigma_{*}\ne0$ exists \cite{SinaiDiffusion,PodolskyEnqvist,PodolskyJokelaMajumder}. The latter determines dynamics of $P(\phi,t)$ at $t\to\infty$, and the diffusion process
is dramatically slowed down --- instead of the linear diffusion law $\langle\phi^{2}\rangle\sim t$
one has weak logarithmic behavior
%%%%%%%%%%%%%%%%%%%%%%
\begin{equation}
\langle\phi^{2}\rangle\sim{\rm log}^{4}t.
\label{eq:LogDependence}
\end{equation}
%%%%%%%%%%%%%%%%%%%%%%

Similar effect is present on the landscape of dS vacua with disorder
\cite{PodolskyJokelaMajumder}, where ``old'' inflation becomes
eternal due to the possibility of multiple consequent Coleman-de Luccia
tunnelings between different de Sitter vacua. Dynamics of the probability
$P_{i}(t)$ for an observer to find herself in a bubble of dS vacuum with the cosmological
constant $\Lambda_{i}$ is described by the vacuum dynamics equations
\cite{VacuumDynamics}
%%%%%%%%%%%%%%%%%%%%%%
\begin{equation}
\frac{dP_{i}}{d\tau}=\sum_{j}\left(\Gamma_{ji}P_{j}-\Gamma_{ij}P_{i}\right)=\sum_{j}H_{ij}P_{j},
\label{eq:VacDyn}
\end{equation}
%%%%%%%%%%%%%%%%%%%%%%
where $\Gamma_{ij}$ are Coleman-de Luccia tunneling rates (inverse
characteristic times of transition) between de Sitter vacua $i$ and
$j$. Technically, equations (\ref{eq:VacDyn}) represent
the discretized version of the multifield Fokker-Planck equation (\ref{eq:fokkerplanck}).
Not surprisingly, it turns out that the case of weakly disordered
tunneling rates \cite{PodolskyJokelaMajumder}
%%%%%%%%%%%%%%%%%%%%%%
\begin{equation}
\Gamma_{ij}=\bar{\Gamma}\delta_{ij}+\delta\Gamma_{ij}
\end{equation}
%%%%%%%%%%%%%%%%%%%%%%
is reduced to the two types of disorder (random potential and random
force) discussed above. The effects of disorder are also similar:
when the number of vacua adjacent to any given one is equal to $2$,
the diffusion of the probability distribution $P_{i}(t)$ is suppressed
as
%%%%%%%%%%%%%%%%%%%%%%
\begin{equation}
\langle n^{2}(t)\rangle\sim{\rm log}^{4}t
\end{equation}
%%%%%%%%%%%%%%%%%%%%%%
at $t\to\infty$ (here $n$ is the index numerating dS vacua on the
landscape).

Let me now discuss how RG analysis of the effective action (\ref{eq:EffectiveAction1})
may help to resolve the gauge invariance problem for eternal inflation.
The probability distribution $P(\phi,t)$ governed by the Eq. (\ref{eq:fokkerplanck})
does not take exponentially inflating 3-dimensional volumes of Hubble
patches into account. The reason for that is the effective coarse
graining \cite{StochasticInflation} at scale $l\sim H^{-1}$ performed
during the derivation of the Fokker-Planck equation (\ref{eq:fokkerplanck}).
Usually, the effect of volume growth is taken into account \cite{VolumeWeightedMeasure} (see also \cite{VWMrecent} for recent developments) by adding the volume weighting term to the r.h.s. of the Fokker-Planck
equation:
%%%%%%%%%%%%%%%%%%%%%%
\begin{equation}
\partial_{t}P_{V}=\partial_{\phi}\left(\frac{V'(\phi)}{3H(\phi)}P_{V}\right)+\frac{1}{8\pi^{2}}
\partial_{\phi}\left(H^{3/2}(\phi)\partial_{\phi}H^{3/2}(\phi)P_{V}\right)+3H(\phi)P_{V}
\label{eq:FPvolumewighted}
\end{equation}
%%%%%%%%%%%%%%%%%%%%%%
Since $H$ is positively defined, patches with larger 3-dimensional
volume and higher $H$ are clearly dynamically favored, which leads
to many technical and conceptual problems associated with the measure
$P_{V}(\phi,t)$. Probably, the most widely discussed one among them
is the \emph{gauge invariance problem }\cite{GaugeInvariance} ---
the probability $P_{V}$ and associated correlation functions of physical
observables are exponentially sensitive to the choice of time reparametrization
\footnote{One may even require stronger invariance of $P_{V}$ w.r.t. reparametrizations
$dt=f(\tau,\phi)d\tau$.}
%%%%%%%%%%%%%%%%%%%%%%
\begin{equation}
dt=f(\tau)d\tau.
\label{eq:TimeReparametrization}
\end{equation}
%%%%%%%%%%%%%%%%%%%%%%
Another important problem, directly related to the previous one is
very strong dependence of the probability distribution $P_{V}$ on
initial conditions for eternal inflation, as can be transparently
seen for probability measures with effective volume cutoff \cite{VolumeCutoff}.
Before I proceed to the discussion of RG analysis and RG fixed points
of the effective action (\ref{eq:EffectiveAction1}), a couple of remarks
has to be made.

First of all, one may observe that there exists a problem with Eq.
(\ref{eq:FPvolumewighted}) even more serious than the problem of
gauge invariance --- the probability flow described by the modified
Fokker-Planck Eq. (\ref{eq:FPvolumewighted}) is \emph{non-unitary},
i.e., the total probability $\int d\phi P(\phi,t)$ is not conserved.
It was understood long time ago \cite{NakaoNambuSasaki} how write
the Fokker-Planck equation for the \emph{normalizable probability}
--- a non-local term proportional to $\langle H\rangle$ has to be
added to the r.h.s. of the Eq. (\ref{eq:FPvolumewighted}):
%%%%%%%%%%%%%%%%%%%%%%
\bea\label{eq:FPcorrected}
\partial_{t}P(\phi,t)&=&\partial_{\phi}\left(\frac{V'(\phi)}{3H(\phi)}P(\phi,t)\right)+\frac{1}{8\pi^{2}}
\partial_{\phi}\left(H^{3/2}(\phi)\partial_{\phi}H^{3/2}(\phi)P(\phi,t)\right)+\\
& & +3P(\phi,t)H(\phi)-3P(\phi,t)\int d\phi'H(\phi')P(\phi')
\eea
%%%%%%%%%%%%%%%%%%%%%%
Dynamics of $P$ described by the Eq. (\ref{eq:FPcorrected}) is much
healthier than the dynamics of $P_{V}$. In particular, the exponential
sensitivity of the probability measure $P$ to the time reparametrizations
(\ref{eq:TimeReparametrization}) is lost, although the issue of gauge
invariance is of course not yet resolved completely.

Second, one actually has to ask herself --- is it true that correlation
functions of \emph{any} physical observable should be invariant w.r.t.
the reparametrizations (\ref{eq:TimeReparametrization})? The answer
to this question is certainly negative, and to illustrate this, the
following counter-example can be presented. Instead of calculating
correlation functions of the inflaton we can calculate correlation
functions of the world time $t=t(\phi)$ or the number of efoldings
$N(\phi)=\log a=\int^{t}Hdt'$. The world time and the number of efoldings
are two different quantities, related to physically different observables:
$t$ is essentially the quantum phase of a heavy ($m\gg H$) particle
living in a given Hubble patch, while $N(\phi)$ is the phase of a
light ($m\ll H$) particle. Clearly, the correlation functions of
these two observables should behave differently, and both observables are not invariant w.r.t. the transformations (\ref{eq:TimeReparametrization}).

With these remarks in mind, let us still try to find whether solutions
of the Eq. (\ref{eq:FPcorrected}) exist which weakly depend on initial
conditions and are invariant w.r.t. the transformations (\ref{eq:TimeReparametrization}).
First of all, we note that volume weighting term $3HP$ and the non-local
term $-3\langle H\rangle P$ cancel each other if time is measured
in units of local Hubble parameter $H$, and one has
%%%%%%%%%%%%%%%%%%%%%%
\begin{equation}
\partial_{\log a}P=\partial_{\phi}\left(\frac{V'(\phi)}{3H(\phi)}P\right)+\frac{1}{8\pi^{2}}\partial_{\phi}\left(H^{3/2}(\phi)
\partial_{\phi}H^{3/2}(\phi)P\right).
\label{eq:ConformalTime}
\end{equation}
%%%%%%%%%%%%%%%%%%%%%%
The general solution of this equation has the form similar to (\ref{eq:expansion}).
In particular, at $\log a\to\infty$ it approaches the Hartle-Hawking
asymptotics (\ref{eq:Asymp}). Of course, the latter a) does not depend
on initial conditions since it is the attractor for the ensemble of trajectories
$P=P(\phi,t)$ and b) is invariant w.r.t. reparametrizations of $t$
simply because the expression (\ref{eq:Asymp}) is the function of
the inflaton potential $V(\phi)$ only.

However, as was explained above, in at least two important classes
of inflationary models this asymptotics is never reached because the
Hartle-Hawking wavefunction is not normalizable: (a) infinite landscape
of vacua a la condensed matter with effective potential/Hawking-Moss/Coleman-de
Luccia tunneling rates not growing at infinity, when transport properties
of the probability flow $P$ become of especial importance, and (b)
models of chaotic inflation where the regime of deterministic slow
roll acts as a sink for the probability flow (\ref{eq:fokkerplanck})
and inflation ends after slow roll conditions break down. Fortunately,
analogues of the Hartle-Hawking asymptotics exist for both classes
of models which have the properties of invariance w.r.t. reparametrizations
of $t$ and independence of initial conditions.

Let me start with the class of models (a), where the hint is provided
by the dynamical renormalization group analysis of the effective action
(\ref{eq:EffectiveAction1}). As we have discussed, long time dynamics
of the probability distribution $P(t)$ is determined by the trivial
RG fixed point corresponding to the regime of linear diffusion $\langle\phi^{2}\rangle\sim t$
unless non-trivial RG fixed points for the effective action
(\ref{eq:EffectiveAction1}) exist. These RG fixed points correspond to
attractors for the ensemble of trajectories $P=P(t)$. Once the probability
distribution reaches such attractor, memory about initial conditions
is lost.\footnote{This is not so for the trivial RG fixed point corresponding to the linear diffusion law $\langle\phi^2\rangle\sim{}t$ --- the gaussian distribution
(\ref{eq:Gaussian}) is always peaked at the same point where the
initial distribution is peaked.} On the other hand, $P(t)$ is invariant w.r.t. time reparametrizations
(\ref{eq:TimeReparametrization}) due to the conformal invariance
of the correlation functions at the RG fixed point. I have presented
above one such example of the inflationary theory with dynamical RG
fixed point: inflation on the low-dimensional landscape with disorder,
when the behavior of the correlation function $\langle\phi^{2}\rangle$
is logarithmic.\footnote{I would also like to mention another well known example of the theory
with the dynamical RG fixed point, which perfectly illustrates why the analysis of RG fixed points is important
--- the theory of Kolmogorov's turbulence. The MSR effective action for the Navier-Stokes (or Euler)
equation in the presence of stochastic noise has always at least two RG fixed points. One of them corresponds to the ultimate equilibrium, $v=0$ and is trivial, while another one is not --- it corresponds to the Kolmogorov's cascade (of decaying turbulence). The latter is the
stationary flow of energy from larger scales towards smaller scales.
In the regime of cascade, correlation functions of observables (such
as velocity of the fluid $v$) are scale invariant with non-trivial
scaling properties, and any information about initial conditions for
these correlation functions is lost as long as the regime of cascade
is reached, since Kolmogorov cascade is the attractor.}

Non-trivial stationary solutions of the Fokker-Planck equation also
exist for chaotic inflationary models (class (b) above) \cite{StochasticInflation}.
To find them, one needs to rewrite the Fokker-Planck equation (\ref{eq:ConformalTime})
in the form
%%%%%%%%%%%%%%%%%%%%%%
\bea\label{eq:current}
\frac{\partial P}{\partial\log a}=-\frac{\partial j}{\partial\phi},& &j=-\frac{1}{3\pi M_{P}^{2}}\partial_{\phi}(VP)-\frac{M_{P}^{2}}{8\pi}\frac{V'}{V}P.
\eea
%%%%%%%%%%%%%%%%%%%%%%
Instead of the Hartle-Hawking wavefunction (\ref{eq:Asymp}) the time
independent asymptotics of the general solution of the Eq. (\ref{eq:current})
is given by the solution with $j={\rm Const.}$ Such solutions describe
constant flow of probability towards the regime where inflation becomes
deterministic (i.e., when the condition (\ref{eq:EternalInfCondition})
breaks down) and ends afterwards. They also have properties similar
to the Hartle-Hawking asymptotics (\ref{eq:Asymp}) --- they are 1)
dynamical attractors and therefore do not carry information about
initial conditions and 2) transparently invariant w.r.t. the time
reparametrizations (\ref{eq:TimeReparametrization}).

\noindent
\bigskip

{\bf Acknowledgments}

\bigskip

I would like to thank A. Starobinsky for numerous discussions and explanations.
This work was partially supported by the EU 6th Framework Marie Curie Research and Training network ``UniverseNet'' (MRTN-CT-2006-035863), RFBR grant 08-02-00923 and ``Leading Scientific Schools grant'' of RAS 4899.2008.2.

\end{document}